\documentclass[twocolumn,preprintnumbers,amsmath,amssymb,aps]{revtex4}

\usepackage{graphicx}% Include figure files
\usepackage{dcolumn}% Align table columns on decimal point
\usepackage{bm}% bold math

\def\beq{\begin{equation}}
\def\eeq{\end{equation}}
\def\be{\begin{equation}}
\def\ee{\end{equation}}
\def\ba{\begin{array}}
\def\ea{\end{array}}
\def\bea{\begin{eqnarray}}
\def\eea{\end{eqnarray}}

\def\eps{\epsilon}

\begin{document}

\title{Propagating spinors on a tetrahedral spacetime lattice}

\author{Brendan Z. Foster}
 \email{bzf@physics.umd.edu}
\altaffiliation[Present address: ]
{Insitut d'Astrophysique de Paris, 98 bis Bvd.~Arago, 75006 Paris,
FRANCE}%Lines break automatically or can be forced with \\
 \affiliation{Department of Physics, University of Maryland\\ College
Park, MD 20742-4111 USA}%Lines break automatically or can be forced with\\
\author{Ted Jacobson}%
 \email{jacobson@physics.umd.edu}
 \altaffiliation[Present address: ]
{Insitut d'Astrophysique de Paris, 98 bis Bvd.~Arago, 75006 Paris,
FRANCE}%Lines break automatically or can be forced with \\
\affiliation{Department of Physics, University of Maryland\\ College Park,
MD 20742-4111 USA}


\begin{abstract}

We derive a discrete path integral for massless fermions
on a hypercubic spacetime lattice with null faces.
The amplitude for a path with $N$ steps and $B$ bends
is $\pm(1/2)^N(i/\sqrt{3})^B$.

\end{abstract}

\maketitle

\newcommand{\amp}{A(\theta)}
\newcommand{\ampk}{A(\epsilon k)}
\newcommand{\n}[1]{\hat{n}_{#1}}
\newcommand{\bra}[1]{|+#1\rangle}
\newcommand{\N}[1]{n_{#1}^{\mu}}
\renewcommand{\d}{\textrm{d}}
\newcommand{\tht}[1]{\theta_{#1}}
\newcommand{\delx}{\triangle x}
\renewcommand{\k}{\mathbf{k}}
\newcommand{\x}{\mathbf{x}}
\def\bfs{\mathbf{\sigma}}
\def\a{\alpha}
\def\D{\Delta}
\def\ra{\rangle}
\def\la{\langle}

In this article, we discuss a four-dimensional variant of
`Feynman's Checkerboard' problem~\cite{qmpi}---a construction of
the propagator for the Dirac equation in 1+1 dimensions via a
sum-over-paths method, where the paths traverse a
null lattice. In that problem, the amplitude for a given zig-zag
path is just $(i\epsilon m)^R$, where $\epsilon$ is the
time duration of a lattice step, $m$ is the particle mass,
and $R$ is the number of direction reversals (and we use units
with $\hbar=c=1$). If the mass vanishes the particle moves only
to the right or the left with the speed of light, and these two
motions correspond to the two chiralities for a
Dirac spinor in 1+1 dimensions.

Feynman's checkerboard path integral is striking in its
simplicity, and intriguing because it accounts for relativistic
propagation and Dirac matrices with nothing more than a 
simple factor of $i$ associated with a geometric 
property---a bend---in a piecewise lightlike
path. 
It hints at the possibility that a simple, discrete dynamics
could underly the usual continuum  description of relativistic 
quantum field theory.
A longstanding question has been whether or not
a path integral could be found in $3+1$ dimensions
preserving these surprising features. Here we show that
one can indeed come very close. 

Most work on lattice
formulations of spinor propagation
is directed at lattice field theory calculations,
and thus involves Grassmann variables and 
``path" integrals over field configurations
in a spacetime of Euclidean signature~\cite{x}.
Here instead we seek a
bosonic (i.e.~non-Grassmanian)
formulation in terms of a sum over particle paths
on a Minkowski signature lattice.
Such formulations have been studied 
previously in~\cite{TJdiss,Bialynicki-Birula:hi} and
references therein. 
What is new here is the
interesting structure of the lattice employed, and the
fact that it allows for a particularly simple rule
for the amplitudes. 

We focus on the massless case, which 
in 3+1 dimensions is far more interesting
than in 1+1 dimensions.
The decoupled right and left chiralities are then described
by two-component spinors satisfying the Weyl equation,
\beq
\sigma^\mu\partial_\mu\Psi=0,
\label{covWeyl}
\eeq
where $\sigma^{\mu}=(1,\pm\vec{\sigma})$
with $\vec{\sigma}$ the Pauli matrices.
(The + sign corresponds to right handed spinors.)
%We discretize spacetime with a
%hypercubical lattice oriented so one
%diagonal of the hypercube lies in the time
%direction, and with the step speed chosen
%three times the speed of light,
%so that the discrete light cone just encloses
%the continuum one. Much as in Feynman's example,
%we attach an amplitude  $\pm i/\sqrt{3}$
%to each bend in the path, where 
%the sign depends upon the order of the steps.
%Together with a factor of one-half for each step,
%the resulting sum-over-paths
%produces the usual retarded propagator for the
%Weyl equation.

%We begin with a description of the lattice, then
%derive the discrete path integral for the retarded
%propagator through the finite
%differencing method of~\cite{Jacobson:xt}. This results
%in a `spinor chain path integral', which
%by suitable choices of the spinor phases
%takes on the form mentioned above.
%We then show that in the limit of infinitesimal lattice spacing
%this path integral converges to the continuum result.

The lattice we
use here is topologically hypercubical
but ``tilted on its corner".  A
quantum network based on
this lattice was called ``hyperdiamond"
and previously studied by Finkelstein and
collaborators\cite{Finkelstein:1996wu}
(see also~\cite{Smith:1995kd}).
We use this name here for the lattice itself.

A diagonal of the hypercube
defines our time axis.
The edges from one
vertex lie in the direction of the
four spacetime vectors
\beq
\N{i}=(1,\a \n{i}).
\label{Ni}
\eeq
The four spatial unit vectors $\n{i}$ point to
the vertices of a tetrahedron, and the step
speed $\a$ is for the moment unspecified.
The $\n{i}$ sum to zero, and the inner product
or angle between any distinct two is the same.
Hence for $i\ne j$ we have
$\n{i}\cdot\n{j}=-1/3$, and the angle is
equal to $\cos^{-1}(-1/3) \approx 109^\circ$.
%
%An explicit expression for the components
%of the $\n{i}$ in
%a particular basis is
%
%\bea\label{COORD}
%\n{1}&=&(1,0,0)\\
%\n{2}&=&\textstyle{\frac{1}{3}}(-1,2\sqrt{2},0)\\
%\n{3}&=&\textstyle{\frac{1}{3}}(-1,-\sqrt{2},\sqrt{6})\\
%\n{4}&=&\textstyle{\frac{1}{3}}(-1,-\sqrt{2},-\sqrt{6}).
%\eea
%

\begin{figure}[htb]
\vbox{ \vskip 8 pt
\centerline{\includegraphics[width=4cm]{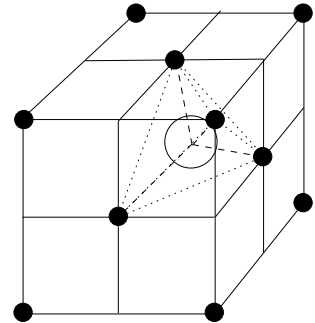}}
\caption{ \label{fcc}
\small Face-centered cubic lattice of points at one time
step. The tetrahedron (dotted lines)
is comprised of the four points reached
from the center of the small cube in one time step
(dashed lines).
The continuum sphere of light is enclosed by and tangent to the
tetrahedron. The distance from the center to a
tetrahedron vertex is three times the radius of the sphere.
The step length $a$ and cube edge length $L $ are shown.
\smallskip}
}\end{figure}

The spatial lattice at one time is a face-centered
cubic (fcc) lattice. A way to see this is to begin with
the tetrahedron of points that lies at
one time step to the future of a given spacetime point $p$. 
The four dimensional lattice has translation
symmetries that map any point to any other point,
and the spatial lattice at one time must share this property.
Hence it can be grown from this tetrahedral seed
by translation along the edges of the tetrahedron,
which produces the fcc lattice shown in Fig.~\ref{fcc}.

Evolving the spatial lattice one time step to the future
amounts to shifting it along the displacement from
the center of one tetrahedron to one of its vertices,
yielding a distinct but equivalent spatial lattice.
After four such steps the original spatial
lattice is recovered.
%One can visualize these steps all in the same
%direction, each one extending for half a diagonal
%of one of the constituent cubes in Fig.~\ref{fcc}.
%Alternatively, one can think of them as successively along
%four distinct edges of the hypercube, which makes it
%obvious that four steps brings one back to the original
%lattice.

%It will be important later to know the spatial volume per point
%at one time step, expressed in terms of the step length.
%It is clear from Fig.~\ref{fcc} that each point can be
%associated uniquely with a pair of constituent
%cubes of edge length $L $.
%The volume of these cubes is $2L ^3$. The step
%length $a$ is half the diagonal of one of these cubes,
%hence $a=\sqrt{3}L /2$, so the volume per point
%$V_p$ is given by
%
%\beq
%V_p= \frac{16}{3\sqrt{3}}a^3.
%\label{Vp}
%\eeq
%

It might seem natural to choose the step speed
$\a=1$, so that the links $\N{i}$ of the hyperdiamond would
be null as envisaged in~\cite{Finkelstein:1996wu}.
However, in this case
the lattice propagator would fail to converge at all in
the continuum limit. The reason is that such a spacetime
lattice violates the well-known ``Courant condition" for stability:
the discrete region of causal influence 
must contain the continuum one.

To marginally satisfy the Courant condition,
the polyhedral cone formed by the
four hypersurfaces spanned by three of the
$\N{i}$'s must be tangent to
the continuum light cone.
That is, the faces of the hypercube must be null.
Thus each of these hyperplanes
must contain one and only one
null direction. By symmetry this null direction must
coincide with the sum of the three link vectors, e.g.
$N^\mu=(3,\a(\n{1}+\n{2}+\n{3}))$. The Minkowski
norm of $N^\mu$ is $9-\a^2$, hence if it is to
be null  we must choose $\a=3$. Moreover,
$N^\mu n_{1\mu}=3-(\a^2/3)$, so if $\a=3$ the null
vector $N^\mu$
is orthogonal to all vectors in the hyperplane,
confirming that the hyperplane is indeed null.

%\section{Spinor chain path integral}

%\emph{Spinor chain path integral}--- Following the method
%of~\cite{Jacobson:xt}, we now make a finite
%difference approximation to the Weyl equation (\ref{covWeyl})
%on the hyperdiamond lattice,
%to arrive at a path integral for the retarded
%propagator in which a path is described by a sequence of
%spinors and the amplitude is the product of inner products
%of successive spinors on the chain.

Now consider the tetrahedral quartet of unit vectors
$\n{i}$. The sums $\sum_i \n{i}^a$ and
$\sum_i\n{i}^{a}\n{i}^{b}$
(with `$\n{i}^{a}$' denoting the `a'component of $\n{i}$)
are invariant under the symmetries of the tetrahedron,
hence the first sum must vanish and the second sum
must be proportional to the Euclidean metric
$\delta^{ab}$. Since the trace is equal to four this yields
the relation
\be
\label{BASIS}
\sum_i\n{i}^{a}\n{i}^{b}=\frac{4}{3}\delta^{ab}.
\ee
Using these identities and the
definition (\ref{Ni}) of the 4-vectors $\N{i}$,
the matrix 4-vector $\sigma^{\mu}=(1,\vec{\sigma})$ can
be expressed as
\be
\sigma^{\mu}=\frac{1}{2}\sum_{i}
\frac{1}{2}\left(1+\frac{3}{\alpha}\n{i}\cdot\vec{\sigma}\right)\N{i}.
\label{sigmaalpha}
\ee
In the special case $\a=3$, for which the polyhedral light
cone conists of null hyperplanes, this becomes just
\be
\sigma^{\mu}=\frac{1}{2}\sum_{i}P_{i}\, \N{i},
\label{sigma}
\ee
where 
\be
P_i={\textstyle\frac{1}{2}}(1+\n{i}\cdot\vec{\sigma})
\label{Projector}
\ee
is the projector for spin up in the direction $\n{i}$.
Using the identity (\ref{sigma})
the Weyl equation (\ref{covWeyl}) for right-handed
spinors takes the form
\beq
\frac{1}{2}\sum_{i}P_{i}\N{i}\partial_\mu\Psi=0.
\label{PWeyl}
\eeq

We consider the hyperdiamond lattice
with step vectors $\epsilon\N{i}$ scaled
by the step size $\epsilon$.
With the partial
derivatives replaced by finite differences,
\be
\epsilon \N{i}\partial_{\mu}\Psi(x)
\cong\Psi(x)-\Psi(x-\epsilon n_{i}),
\ee
the Weyl equation (\ref{PWeyl})
yields the one-step evolution prescription
for determining $\Psi(x)$ on the lattice from
the values $\Psi(x-\epsilon n_{i})$
at the immediately preceding points,
\be
\Psi(x)=\frac{1}{2}\sum_{i}P_{i}\Psi(x-\epsilon n_{i}).
\label{onestep}
\ee

The finite difference equation (\ref{onestep}) 
is different from any typically used in lattice field theory 
calculations.  For a plane wave 
solution of the form $\Psi(x^i) = \exp(i k_\mu x^\mu) \Psi_0$
(with $\Psi_0$ is a constant spinor) it
implies the dispersion equation
\beq
\Psi_0 = \frac{1}{2}\sum_i P_i  e^{-i\eps k_\mu n_i^\mu} \Psi_0.
\label{nodoubles}
\eeq
Expanding in $\eps$ and using equation (\ref{sigma}),
we find at first order in $\eps$ the standard
Weyl equation for momentum eigenstates, 
$\sigma^\mu k_\mu \Psi_0=0$.
The solutions obey the relativistic
dispersion relation, $k_\mu k^\mu=0$,
and the corresponding zero frequency solution
has vanishing wave vector. 

Fermion
doubling does not occur. That is, there are
no additional zero-frequency solutions to Eq.~(\ref{nodoubles}), 
as we argue later. The Nielsen-Ninomiya theorem~\cite{x}
is presumably evaded since this finite difference equation 
does not satisfy the hermiticity condition.
The attendant lack of unitarity
would be a problem for a lattice field theory
application, but it does not present a problem for
our purpose, which is just to extract a representation
of the propagator as the continuum limit of
a sum over paths on the lattice.

Iterating Eq.~(\ref{onestep}) backwards in time, we see that
the retarded propagator between two points can be
written as a sum over multi-step paths involving
moves in the
directions ${\n{i_{1}}\dots\n{i_{N}}}$ at step speed $3c$,
with the
amplitude for such a path given by the operator
\be
\label{OPER}
\frac{1}{2^N}P_{i_N}\dots P_{i_1}.
\ee
The propagator is then the sum of these operators
over all paths that connect the points.
%Inserting the outer product form (\ref{Pi}) for
%the projectors the amplitude (\ref{OPER})
%becomes
In terms of the unit eigenspinor $|i\ra$ of 
$P_i=|i\ra\la i|$ the amplitude (\ref{OPER})
becomes
\be
\label{spinamp}
\frac{1}{2^N}|i_{N}\ra \la i_N|i_{N-1}\ra\la
i_{N-1}|\cdots|i_2\ra\la i_2|i_1\ra\la i_1|.
\ee
Note that for fixed initial and final points
the number of each of the four step types is fixed, so
one sums only over the order in which the steps are taken.

The amplitude (\ref{spinamp}) is independent of the choice of phases
for the spinors, so we are free to adjust those phases to produce a
particularly nice result. The
unit spinor corresponding to a unit vector with spherical angles
$(\theta,\phi)$ is $(\cos(\theta/2),\sin(\theta/2)\exp(i\phi))$
times an arbitrary overall phase.
The spinors $|i\ra$, which
correspond to the four unit vectors pointing to
the vertices of a tetrahedron, can be taken as
\bea
|1\ra&=&e^{i\psi_{1}}\left(\ba{c}1\\0\ea\right)\hfill\hspace{1.2cm}
|2\ra=\frac{e^{i\psi_{2}}}{\sqrt{3}}\left(\ba{c}1\\
\sqrt{2}\ea\right)\bigskip\nonumber\\
|3\ra&=&\frac{e^{i\psi_{3}}}{\sqrt{3}}\left(\ba{c}1\\
\sqrt{2}\,e^{i\frac{2\pi}{3}}\ea\right)~
|4\ra=\frac{e^{i\psi_{4}}}{\sqrt{3}}\left(\ba{c}1\\
\sqrt{2}\,e^{i\frac{4\pi}{3}}\ea\right)
\label{tspinors}
\eea
If we choose $\psi_1=-\pi/2$ and $\psi_2=\psi_3=\psi_4=0$,
then the inner products between
the various spinors become identical up to a sign that
depends on the order:
\be\label{PHASE}
\la1|2,3,4\ra=\la2|3\ra=\la3|4\ra=\la4|2\ra=\frac{i}{\sqrt{3}}.
\ee
Thus the
amplitude for an $N$ step path with $B$ bends is
\be
\pm\frac{1}{2^N}\left(\frac{i}{\sqrt{3}}\right)^{B}.
\label{amplitude}
\ee
To eliminate the operator $|i_N\ra\la i_1|$ from the amplitude
(\ref{spinamp}), we considered here the matrix element of the propagator
with respect to fixed inital and final spinors $|i_0\ra$ and
$|i_{N+1}\ra$ selected from the set (\ref{tspinors}).
These specify directions of arrival
at the initial point and departure from the final point,
and the amplitude (\ref{amplitude}) includes any contributions
from initial and final bends.
Unfortunately we have not been able to find
a simple way to specify the overall sign other than by reference
to the positive two-step orders given in (\ref{PHASE}).

For left handed Weyl spinors the amplitude for a step
is given by the orthogonal spin projector relative to the
right handed case. Hence we can use the charge conjugates
of the four spinors (\ref{tspinors}), with the result
that the imaginary unit $i$ in the amplitude (\ref{amplitude})
is replaced by $-i$. The effect of a mass $m$ can be included
by allowing for chirality flips between right and left 
handed spinor propagation at each time step, 
with an associated amplitude factor 
$i\epsilon m$~\cite{Jacobson:xt}, 
as on Feynman's checkerboard.

%\emph{The continuum limit}--- 
We now evaluate the sum over paths
for the lattice propagator
${K}_{\epsilon}(\D x)$
for a spacetime displacement
$\D x$
and demonstrate
that it reproduces the continuum propagator in the
limit $\epsilon\rightarrow0$.

The lattice displacement $\D x$ can be expanded
in terms of the four basis vectors:
\beq
\D x^\mu = \sum_j\D x^j \N{j},
\label{Dxi}
\eeq
hence the displacement is determined by a
unique set of four integers $N^j=\D x^j/\epsilon$.
The constraint that a path
connects the two points of interest can be incorporated
as four Kronecker deltas, which we express in a Fourier
representation:
\be\prod_j\delta(N^j, {\D x^j}/{\epsilon})
=\int_{-\pi}^{\pi}
\frac{d^4\theta_j}{(2\pi)^4}
e^{i\sum_j\theta_j (N^j-{\D x^j}/{\epsilon})}.
\ee
The lattice propagator is given by
\be
{K}_{\epsilon}(\D x) =
\sum_{N=0}^{\infty}\int_{-\pi}^{\pi}
\frac{d^4\theta_j}{(2\pi)^4}e^{-i\sum_j\theta_j\D
x^j/{\epsilon}}[\amp]^{N},
\label{PROP}
\ee
where
\beq
\amp=\frac{1}{2}\sum_jP_j\, e^{i\theta_j}.
\label{A}
\eeq
The sum
over $N$ of $[\amp]^N$ produces every possible sequence of
projection operators, each with the appropriate exponential
factor encoding the number of steps in each direction.
When the integrals over $\theta_j$ are carried out,
only those step sequences that produce the displacement
$\D x^\mu$ will survive. In particular, only the
value of $N$ equal to the total number of steps
contributes.

As the step size $\epsilon$ goes to zero, the number of steps
$N$ for a fixed time interval
goes to infinity as $\Delta t/\epsilon$. Convergence
thus requires that the norm $\|\amp\|$
(i.e. the maximum norm of $\amp$ acting on a unit spinor)
be less than or equal to unity. 
%When all the $\theta_j$
%vanish, $\amp$ is just the identity operator, so that condition
%is met. 
%We have shown by a tedious method (although it seems
%a simple demonstration should exist)
%It can be shown~\cite{arxivversion}
It is shown in the appendix
that $\|\amp\|<1$ except when 
at least three of the $\theta_j$ coincide.
Thus, as $N$ becomes larger, $[\amp]^N$ 
converges to zero pointwise except at 
these degenerate $\theta_j$ values.
(Had we assumed an arbitrary step-speed
$\alpha$, convergence would have required $\alpha\ge3$.)

When three $\theta_i$'s coincide
$\amp$ takes the form
$\amp=\frac{1}{2}(e^{i\theta_1}+e^{i\theta_2}) P_1 
+ e^{i\theta_2} \tilde{P}_1$,
where $\tilde{P}_1$ is the projector orthogonal to $P_1$.
It is easily seen that the action of $\amp$ then decreases
the norm of any spinor except an eigenspinor of $P_1$.
[In the exceptional case where 
all four $\theta_i$ coincide $\amp$ is just 
$e^{i\theta}$ times the identity. This leads to no 
difficulty,
as the integral over $\theta=\sum_i\theta_i$ in (\ref{PROP}) 
just produces a Kroneker delta which sets $N$ equal to
the total number of steps.]
Fermion doubling would occur if Eq.~(\ref{nodoubles})
had an extra solution with zero frequency. This would
be an eigenvector of $\amp$ with unit eigenvalue, where
$\theta_i=-\epsilon k_\mu n_i^\mu$ and $\sum_i\theta_i=0$.
Inspection of the above degenerate form of $\amp$ reveals 
that such an eigenvector exists only if all $\theta_i$ vanish,
i.e. if $k_\mu=0$.

%It is the demand for convergence that
%reqires the step-speed to
%be at least three times the speed of light as mentioned
%earlier. Had we assumed an arbitrary step-speed
%$\alpha$, in place of $\amp$ (\ref{sigmaalpha}) would have yielded
%instead
%
%\be
%{1\over{4}}\sum_j\left(1+{3\over{\alpha}}
%\n{i}\cdot\vec\sigma\right)e^{i\theta_j}.
%\ee
%
%When all
%$\theta_j$ are zero, $\amp$ is the identity; however,
%if for example $\theta_1=\delta$ is very small while the
%rest vanish then we have
%
%\be
%A(\delta,0,0,0)=
%1+{1\over{4}}(1+\frac{3}{\alpha}\n{1}\cdot\vec\sigma)(i\delta -
%{1\over{2}}\delta^{2}+O(\delta^{3})).
%\label{Adelta}
%\ee
%
%Since the eigenvalues of $\n{1}\cdot\vec\sigma$ are $\pm1$,
%the squared magnitudes of the eigenvalues of (\ref{Adelta})
%are
%
%\be
%|\lambda_{\pm}|^{2}=
%1-\frac{\delta^{2}}{16}(1\pm\frac{3}{\alpha})(3\mp\frac{3}{\alpha})
%+O(\delta^{3}).
%\ee
%
%For convergence, these values must not be greater than one,
%hence $\alpha$ must not be less than three.

%Returning now to the $\alpha =3$ lattice,
We next introduce the new variables
$k_j:=\theta_j/\epsilon$, in terms of which (\ref{PROP}) takes the form
\be
{K}_{\epsilon}(\D x) =
\epsilon^4\sum_{N=0}^{\infty}\int_{-\pi/\epsilon}^{\pi/\epsilon}
\frac{d^4k_j}{2\pi}e^{-i\sum_jk_j\D
x^j}[\ampk]^{N}.
\label{PROPk}
\ee
Taking the limit $\epsilon\rightarrow0$, 
with the time interval $\Delta t=N\epsilon$ fixed, we have
\bea
[\ampk]^N&=& \left[\sum_j \frac{1}{2}P_j e^{i\epsilon k_j}\right]^N\\
%&=& \left[\sum_j \frac{1}{2}P_j \bigl(1+i\epsilon k_j+O(\epsilon^2)\bigr)\right]^N\\
%&=&  \left[1 + \sum_j \frac{i}{2}\epsilon k_j P_j +O(\epsilon^2)\right]^N\\
&=& e^{iN\epsilon\sum_j  k_j P_j/2}+O(N\epsilon^2).\label{Nepsquared}
%&=& \exp\left[ \frac{i}{4}N\epsilon\sum_j k_j(1+\n{i}^a\sigma^a)\right]+O(N\epsilon^2).
\eea
Moreover the limits of integration in (\ref{PROPk}) approach $\pm\infty$,
hence (\ref{PROPk}) yields
\be
{K}_{\epsilon\rightarrow0}(\D x) =
\epsilon^4\sum_{N=0}^{\infty}\int_{-\infty}^{\infty}
\frac{d^4k_j}{2\pi}e^{-i\sum_jk_j\D x^j}
e^{iN\epsilon\sum_j  k_jP_j/2}
\label{PROPk0}
\ee

The components $\D x^j$ are defined in (\ref{Dxi})
relative to the tetrahedral basis of 4-vectors $\N{i}$.
The quantities $k_j$ can be viewed as components
of a co-vector in the dual basis, and the corresponding
components in an arbitrary basis are denoted $k_\mu$,
\beq
k_j=k_\mu n_j^\mu.
\label{ki}
\eeq
Hence we have $\sum_j k_j\D x^j = k_\mu \D x^\mu$.
Substituting (\ref{ki}) for $k_j$, and using (\ref{sigma}),
the sum in the last exponent of (\ref{PROPk0}) becomes
$k_\mu\sigma^\mu$.
Changing integration variables from $k_j$ to $k_\mu$ in
(\ref{PROPk0}) gives rise to a Jacobian
$|{\partial k_j}/{\partial k_\mu}|=|n_j^\mu|$, which can be
computed using an explicit form of the tetrad
of unit 3-vectors, yielding
 \be
\d^4k_j = 48\sqrt{3}\,\d^4 k^\mu.
\ee

The final step in taking the limit is to replace the discrete variable
$N$ by a continuous one $s=N\epsilon$, in terms of which
the sum $\sum_N$ becomes $\int ds/\epsilon$. With this replacement,
and the change of variables from $k_j$ to $k_\mu$,
(\ref{PROPk0}) becomes
\be
{K}_{\epsilon\rightarrow0}(\D x) =
48\sqrt{3}\epsilon^3\int_0^\infty ds\int_{-\infty}^{\infty}
\frac{d^4k}{(2\pi)^4}e^{-ik_{\mu}\D x^{\mu}}
e^{is k_\mu\sigma^\mu}
\label{PROPs}
\ee
Except for the peculiar factor in front, this is just the
continuum retarded propagator, albeit in a perhaps slightly
unfamiliar form. The integral over $k_0$ produces a Dirac
delta function $\delta(s-\D x^0)$, after which the $s$-integral
sets $s$ equal to $\D x^0$ (assuming $\D x^0>0$),
yielding the retarded propagator in the
more common guise of a three dimensional Fourier transform.
%over $s$ produces (assuming $\D x^0>0$)
%
%\be
%{K}_{\epsilon\rightarrow0}(\D x) =
%48\sqrt{3}\epsilon^3\int_{-\infty}^{\infty}
%{d^3k\over{(2\pi)^3}}e^{-ik_{a}\D x^{a}}
%e^{i\D x^0 k_a\sigma^a}.
%\label{PROP3}
%\ee
%
%This is proportional to the retarded propagator in the
%usual form of a three dimensional Fourier transform.
%Both (\ref{PROPs}) and (\ref{PROP3}) are easily
%seen to satisfy the homgeneous Weyl equation, except at
%$\D x^0=0$ where there is delta function source term.
Had we kept the subleading terms of order
$N\epsilon^2$ ($=s\epsilon$) in (\ref{Nepsquared}) the
convergence factor for the integration
limit  $s\rightarrow\infty$ would have been supplied
much as in~\cite{Jacobson:xt}.

%To derive the latter directly, write the Weyl equation
%(\ref{Weyl}) in
%the form $i\\partial_t\Psi=H\Psi$, with
%the Hamiltonian $H=\vec\sigma\cdot\vec{p}$.
%The retarded propagator is just the matrix elements
%of the evolution operator $\exp(-iHt)$,
%
%\bea
%\label{PSI}\nonumber
%\Psi(\mathbf{x},t)&=&\int\d^3
%x'\langle\mathbf{x}|e^{i\mathbf{p}\cdot\vec\sigma(t-t')}|\mathbf{
%x'}\rangle\la \mathbf{x'} \Psi(t')\\
%&=&\int \d^3 x'\int_{-\infty}^\infty{\d^3
%k\over(2\pi)^3}\,e^{i\mathbf{k}\cdot\vec\sigma(t-t')}
%e^{-i\mathbf{k}\cdot(\mathbf{x}-\mathbf{x'})}\Psi(\mathbf{x'},t'),
%\eea
%
%where the second line can be obtained by inserting a complete
%set of momentum eigenstates...

It remains to account for the prefactor $48\sqrt{3}\, \epsilon^3$.
We computed the propagator
to go between two points on the hyperdiamond lattice. In the continuum,
the amplitude to arrive at one point starting from another
point is zero, since only by integrating
over a finite region should a nonzero amplitude arise.
The prefactor is none other
than the volume per point 
%(\ref{Vp}) 
in the lattice,
with the step length $a$ equal to $3\epsilon$. 
Hence what we have actually
obtained is the continuum propagator integrated over the volume
associated with the initial lattice point.

We have shown that a very simple discrete path
integral converges
in the continuum limit to the retarded propagator for the Weyl equation.
Although the underlying lattice is not Lorentz invariant, that
symmetry is recovered by the propagator in the continuum limit.

Another symmetry not possessed by the discrete propagator is unitarity.
This is not because discreteness and unitarity are necessarily in
conflict.
Indeed, as shown in \cite{Bialynicki-Birula:hi},
one can write a unitary discrete
evolution rule on a body centered cubic lattice (or on an alternating
pair of simple cubic lattices) whose continuum limit is the Weyl equation.
(Interestingly, unitarity and locality were shown there to {\it imply} the
Weyl equation.) 
Consider an
initial state that is non-vanishing only at one lattice point, with
normalized spin state $|\psi\ra$.
At the next time step according to (\ref{onestep})
it has support at the four corners of a tetrahedron, with the amplitudes
$\frac{1}{2}P_i|\psi\ra$. The norm of the state after one step is then
$\la\psi|\sum_i \frac{1}{4}P_i|\psi\ra=\frac{1}{2}$, i.e. it has
decreased by a factor of two, violating unitarity.

It is not just the norm change that
violates unitarity. Also, the evolutions of orthogonal states
do not remain orthogonal. 
Two points at one time
have either one or no common points in their one-step future.
In the former case, the one-step evolutions of two orthogonal
states concentrated
on the two initial points are clearly not orthogonal,
because they overlap in just one point which will
make the unique non-zero (since $P_iP_j\ne0$)
contribution to the inner product.
%This means that initially disjoint 
The
evolutions
therefore
have ``more overlap than they should", which
presumably counteracts the loss of norm of each individual evolution in
such
a way that the continuum limit is unitary.

\section*{Acknowledgements}
We are grateful to E.~Hawkins for
mathematical aid.
This work was supported in part by the NSF under grants
PHY-9800967 and PHY-0300710 at the University of Maryland.

\appendix*
\section{}

In this appendix we prove that the norm of the matrix $A$ defined in 
Eq. (\ref{A}) is less than unity unless at least three of the
$\theta_i$ coincide, in which case the norm is unity. The proof
is due to Eli Hawkins.

Let $|\nu\ra$ be any unit spinor. The squared norm of $A|\nu\ra$ is 
$\|A|\nu\ra\|^2=\la\nu|A^\dagger A|\nu\ra$, whose maximum
is the larger eigenvalue of $A^\dag A$.  
This value defines the squared norm $\|A\|^2$.

Using the definition of the spin projection operators
(\ref{Projector}) and the inner products of the unit vectors
$\hat{n}_i$ ($1$ if $i=j$ and $-\frac13$ if $i\ne j$) we find 
\beq
{\rm tr} (A^\dagger A) = 1 + \tfrac16 \sum_{(ij)} \cos(\theta_i-\theta_j)
\eeq
where the sum is over the 6 choices of $\{i,j\}\subset \{1,2,3,4\}$. 
This trace is at most $2$, and therefore the smaller eigenvalue of 
$A^\dag A$ is less than $1$ unless $A^\dag A=1$.
Hence
\beq
\Phi := \det \left(A^\dag A -1\right)
\eeq
has the same sign as $1 - \| {A}\| $.

The matrix $A^\dag A$ is linear in terms of
$e^{i(\theta_i-\theta_j)}$, therefore $\Phi$ is quadratic. Because
$\Phi$ is invariant under all permutations of the $\theta$'s, it can
be written as a quadratic function of the cosines
$\cos(\theta_i-\theta_j)$. Because $\Phi$ vanishes when the $\theta$'s
are all equal, it is convenient to write it in terms of the cosines
minus 1. It thus takes the form,
\bea
\label{Phi1}
\Phi &=& a \sum_{(ij)} \left(1 - \cos[\theta_i-\theta_j]\right) 
+ b \sum_{(ij)} \left(1 - \cos[\theta_i-\theta_j]\right)^2 \nonumber\\ 
&&+  c\!\!\sum_{(ij)(kl)} \left(1 - \cos[\theta_i-\theta_j]\right)
\left(1 - \cos[\theta_k-\theta_l]\right) .
\eea
The last sum is over the 3 partitions of $\{1,2,3,4\}$ into pairs. 
By setting $\theta_1=\theta_2=\theta_3$ in this expression, 
we see that $a=b=0$.
To determine the value of $c$, consider the case that 
$\theta_1=\theta_2=0$ and $\theta_3=\theta_4=\pi$. 
Then $A$ is the hermitian matrix 
$\frac12(\hat{n}_1+\hat{n}_2)\cdot\vec{\sigma}$,
which has eigenvalues $\pm 1/{\sqrt3}$, 
hence $\Phi=4/9$. 
The last sum in \eqref{Phi1} is $8$, so $c=1/18$.

The determinant $\Phi$ is thus given by 
\beq
\label{Phi2}
\Phi = \tfrac1{18}\sum_{(ij)(kl)} 
\left(1 - \cos[\theta_i-\theta_j]\right)
\left(1 - \cos[\theta_k-\theta_l]\right).
\eeq
This satisfies $\Phi\geq 0$, therefore $\|A\|\leq 1$. Each term is
non-negative, therefore $\Phi=0$ only if every term vanishes. 
This occurs only if at least three of the $\theta$'s are equal.

\end{document}